\documentclass[12pt,a4paper,final]{iopart}

\usepackage{iopams}  
\usepackage{graphicx}
\usepackage[breaklinks=true,colorlinks=true,linkcolor=blue,urlcolor=blue,citecolor=blue]{hyperref}
\usepackage{braket}
\usepackage{xcolor}
\usepackage{sidecap}
\usepackage{mathrsfs}
\newcommand{\pvec}[1]{\vec{#1}\mkern2mu\vphantom{#1}}

\begin{document}

\title[]{Nonlocal vs Local Pseudopotentials Affect Kinetic Energy Kernels in Orbital-Free DFT}

\author{Zhandos A.~Moldabekov}
\address{Center for Advanced Systems Understanding (CASUS), 03581 G\"orlitz, Germany}
\address{Helmholtz-Zentrum Dresden-Rossendorf (HZDR), 01328 Dresden, Germany}
\ead{z.moldabekov@hzdr.de}

\author{Xuecheng Shao}
\address{Key Laboratory of Material Simulation Methods \& Software of Ministry of Education, College of Physics, Jilin University, Changchun 130012, PR China}

\author{Michele Pavanello}
\address{Department of Chemistry, Rutgers University, 73 Warren St., Newark, NJ 07102, USA}
\address{Department of Physics, Rutgers University, 101 Warren St., Newark, NJ 07102, USA}

\author{Jan Vorberger}
\address{Helmholtz-Zentrum Dresden-Rossendorf (HZDR), 01328 Dresden, Germany}

\author{Tobias Dornheim}
\address{Center for Advanced Systems Understanding (CASUS), 03581 G\"orlitz, Germany}
\address{Helmholtz-Zentrum Dresden-Rossendorf (HZDR), 01328 Dresden, Germany}

\begin{abstract}
The kinetic energy (KE) kernel, which is defined as the second order functional derivative of the KE functional with respect to density, is the key ingredient to the construction of KE models for orbital free density functional theory (OFDFT) applications.
For solids, the KE kernel is usually approximated using the uniform electron gas (UEG) model or the UEG-with-gap model.
These kernels do not have information about the effects from the core electrons since there are no orbitals
for the projection on nonlocal pseudopotentials. 
To illuminate this aspect, we provide a methodology for computing the KE kernel from Kohn-Sham DFT and apply it to the valence electrons in bulk aluminum (Al) with a face-centered cubic lattice and in bulk silicon (Si) in a semiconducting crystal diamond state. We find that bulk-derived local pseudopotentials provide accurate results for the KE kernel in the interstitial region. 
The effect of using nonlocal pseudopotentials manifests at short wavelengths, defined by the diameter of an ion surrounded by its core electrons.
Specifically, we find that the utilization of nonlocal pseudopotentials leads to significant deviations in the KE kernel from the von Weizs\"acker result in this region, which is, as a rule, explicitly enforced in most widely used KE functional approximations for OFDFT simulations. 

\end{abstract}
\vspace{2pc}

\section{Introduction\label{sec:introduction}}

First principle electronic structure simulations on mesoscopic scales are of paramount importance for material science and related fields~\cite{MADDEN20069}.
On the nanoscale, the work horse is given by Kohn Sham density functional theory (KSDFT)~\cite{Kohn_Sham} due to its often cited balance between accuracy and computational efficiency~\cite{Jones_RMP_2015}.
Indeed, the combination of a KSDFT description of the electrons with a molecular dynamics propagation of the heavier nuclei within the widely used Born Oppenheimer approximation has emerged as a standard tool in a gamut of research fields~\cite{Jones_RMP_2015}.
A popular approach for the description of larger systems for which an explicit DFT solution is no longer feasible is given by so-called \emph{force fields}~\cite{Unke2021}. These are constructed on the basis of \emph{ab initio} DFT results for relatively small systems, and subsequently allow for highly efficient MD simulations of up to $N\sim10^9$ ions or nuclei~\cite{THOMPSON2022108171}.


While being an important step in the right direction,
some applications explicitly require information about electronic density, e.g., to model systems under external static or dynamic perturbations~\cite{Jiang_prb_2022, Krishtal_jcp_2015, Gawne_prb_2024, Moldabekov_PRR_2024, Moldabekov_ACSomega_2024};
such information is unavailable from force-field based methodologies by necessity.
A potential alternative is given by the orbital-free formulation of density functional theory (OFDFT), which is computationally cheaper than KSDFT and allows for the modeling of both ionic dynamics and of the electronic density (see Refs.~\cite{Wenhui_ChemRev_2023, Qiang_WIRE_2024} for topical review articles).
Due to its reduced computational cost~\cite{PRR_2022}, the OFDFT method facilitates simulations of large systems with hundreds of thousands of atoms both at ambient \cite{dftpy_paper} as well as at extreme conditions \cite{Dragon_paper}.

While both methods are formulated from first principles, there is a key difference between KSDFT and OFDFT. In the former method, it is straightforward to evaluate the kinetic energy (KE) $T_s[n]$ of the noninteracting system from the set of Kohn-Sham orbitals. In OFDFT, the absence of orbitals greatly reduces the required computational cost, but the noninteracting KE has to be supplied as an external input. In practice, it has to be approximated.
This has been a topic of active research for decades, with a plethora of various KE functionals being developed and tested over this time \cite{Wenhui_ChemRev_2023, Qiang_WIRE_2024}.  

Another drawback of OFDFT that limits its accuracy compared to KSDFT  is the treatment of core electrons.  In KSDFT, the computationally demanding task of explicitly calculating core states is circumvented using pseudopotentials \cite{Hamann_PRL_1979, Martin_2004}. In fact, the utilisation of pseudopotentials is indispensable to remove the computational bottleneck of all-electron calculations without the loss of accuracy. 
Accurate and transferable nonlocal pseudopotentials (NLPs) are constructed by enforcing various physical constraints on pseudo valence orbitals, such as norm-conservation, core charge reproduction, and the correct scattering behavior. Formally, the nonlocality of a pseudopotential is expressed by the projection of the pseudo-valence orbitals on different angular momentum-dependent potentials for the calculation of electron-ion energies. 

Due to the absence of orbitals, 
OFDFT simulations are usually limited to the utilisation of local pseudopotentials (LPs).

Recently, Xu \textit{et al.} \cite{Xu2022} developed a scheme to entangle NLPs with the KE density functional for the calculation of the electron-ion interaction energy. This is done by projecting a NLP onto the non-interacting density matrix, which is being approximated by a Gaussian. 
Besides the electron-ion interaction, the electronic KE is also directly impacted by the type of pseudopotential used within the self-consistent KS-DFT modeling. Therefore, for OFDFT, it raises the question: What is the impact of using a nonlocal pseudopotential in the kinetic energy density functional compared to a standard local pseudopotential?
In this work, we develop and demonstrate a method for the computation of the KE kernel from KSDFT that explicitly incorporates the effect of NLPs,
 where the KE kernel is defined as the second-order functional derivative of the KE functional; it is the key ingredient for the construction of both semi-local and nonlocal KE functionals.

To demonstrate our methodology, we perform a set of KSDFT calculations using NLPs and bulk-derived local pseudopotentials (BLPs) \cite{B810407G} for aluminum (Al) with a face-centered cubic (fcc) lattice and silicon (Si) in the semiconducting crystal diamond (cd) phase.  The KSDFT results are further compared with the results from OFDFT simulations and with the KE kernels computed using uniform electron gas (UEG; the archetypical system of interacting electrons in a homogeneous background~\cite{quantum_theory,review,loos}) and UEG-with-gap models~\cite{Levine_prb_1982, Constantin_prb_2017}. Our comparative analysis of the KE kernels shows that the effect of using NLPs is most pronounced for density inhomogeneities on length scales smaller than the diameter of the ion together with the core electrons. By comparing the results computed using NLPs and BLPs, we also demonstrate that local pseudopotentials provide accurate data for the KE kernel in the interstitial region of the considered materials.
 
The paper is organized as follows: In Sec. \ref{sec:theory}, we show the connection between the KE kernel and the KS response function and provide the statement of the problem considered in this work. 
In Sec. \ref{sec:theory2},  we introduce the theoretical framework of the external harmonic perturbation method for the calculation of the static KS response function, and we provide details of commonly used approximations for the KE kernel for metals and semiconductors.  The calculation details are provided in Sec. \ref{sec:sim_details}. In Sec. \ref{sec:results}, the KE kernel results for bulk Al with an fcc lattice and for semiconducting bulk cd Si are presented.
We conclude the paper by summarizing our findings and providing an outlook on potential future works.

\section{Theoretical background and statement of the problem\label{sec:theory}}

Throughout this work, we consider a system of electrons at ambient conditions, i.e., at $T=0$. The generalization of our results to finite temperatures as they are relevant e.g.~for the description of exotic \emph{warm dense matter}~\cite{wdm_book,Dornheim_review,new_POP} that occurs in astrophysical objects, laser-excited solids, and inertial fusion energy applications is straightforward, see Sec.~\ref{sec:summary}.

\subsection{KE kernel and KS response function}

The foundation of OFDFT is the energy functional of the electron density, $n(\vec r)$,
\begin{equation}
E[n]  = T_s[n] + E_H[n] + E_{xc}[n] + V_{\rm ext}[n].
\end{equation}
Where $T_s$ is the noninteracting kinetic energy functional, $E_H$ the classical electron-electron repulsion, $E_{xc}$ the exchange-correlation (XC) functional and $V_{ext}$ the interaction with the external potential.

The above leads to the Euler-Lagrange equation derived from the variational minimization of the energy functional as a function of density at a given number of electrons \cite{Wenhui_ChemRev_2023}:
\begin{equation}\label{eq:EL}
    \frac{\delta T_s[n]}{\delta n(\vec r)}+\frac{\delta E_H[n]}{\delta n(\vec r)}+\frac{\delta E_{\rm xc}[n]}{\delta n(\vec r)}+\frac{\delta V_{\rm ext}[n]}{\delta n(\vec r)}=\mu,
\end{equation}
where $\mu$ is the Lagrange multiplier (the electron chemical potential) introduced to keep the density normalized to the number of electrons in simulations. It is worth noting that the external potential is simply given by $\frac{\delta V_{\rm ext}[n]}{\delta n(\vec r)} = v_{ext}(\vec r)$.

The sum of the last three terms on the l.h.s.\ of Eq. (\ref{eq:EL}) defines the KS potential:
\begin{equation}\label{eq:vks}
    v_{\rm KS}[n](\vec r)=\frac{\delta E_H[n]}{\delta n(\vec r)}+\frac{\delta E_{\rm xc}[n]}{\delta n(\vec r)}+\frac{\delta V_{\rm ext}[n]}{\delta n(\vec r)}.
\end{equation}

Taking the functional derivative of Eq. (\ref{eq:EL}) and using definition (\ref{eq:vks}), one finds \cite{Wang_Carter_book}:
\begin{equation}\label{eq:EL20}
    \frac{\delta^2 T_s[n]}{\delta n(\vec r)\delta n(\pvec r^{\prime})}=-\frac{\delta v_{\rm KS}[n](\vec r)}{\delta n(\pvec r^{\prime})}.
\end{equation}

On the other hand, for an equilibrium system with the density distribution $n_{\rm eq}(\vec r)$, the functional Taylor expansion for the change in the KS potential $\Delta v_{\rm KS}(\vec r)$ due to a perturbation in the density yields in first order:
\begin{equation}\label{eq:chiks3}
    \Delta v_{\rm KS}(\vec r)=\int \left. \frac{\delta v_{\rm KS}[n](\vec r)}{\delta n(\pvec r^{\prime})}\right|_{n=n_{\rm eq}} \delta n(\vec r^{\prime}) \mathrm{d}\pvec r^{\prime}.
\end{equation}

Substituting Eq. (\ref{eq:EL20}) into Eq. (\ref{eq:chiks3}), for a system in equilibrium, we find

\begin{equation}\label{eq:chiks4}
    \Delta v_{\rm KS}(\vec r)=\int \left. \frac{\delta^2 T_s[n]}{\delta n(\vec r)\delta n(\pvec r^{\prime})}\right|_{n=n_{\rm eq}} \delta n(\pvec r^{\prime}) \mathrm{d}\pvec r^{\prime}.
\end{equation}

Within linear response theory, the change in the density induced by the variation of the  KS potential is defined by the KS response function $\chi_{\rm KS}(\vec r, \pvec r^{\prime}) $:
\begin{equation}\label{eq:chiks}
    \Delta n(\vec r)=\int \chi_{\rm KS}(\vec r, \pvec r^{\prime}) \Delta v_{\rm KS}[n](\pvec r^{\prime}) \mathrm{d} \pvec r^{\prime}.
\end{equation}

The inverse of the KS response function $\left[\chi_{\rm KS}(\vec r, \pvec r^{\prime})\right]^{-1}$ is introduced by the inversion of Eq. (\ref{eq:chiks}):
\begin{equation}\label{eq:chiks2}
    \Delta v_{\rm KS}(\vec r)=\int \left[\chi_{\rm KS}(\vec r, \pvec r^{\prime})\right]^{-1} \Delta n(\pvec r^{\prime}) \mathrm{d} \pvec r^{\prime}.
\end{equation}

By comparing Eqs. (\ref{eq:chiks2}) and (\ref{eq:chiks4}), we deduce for KE kernel:
\begin{equation}\label{eq:EL2}
    K_{\rm s}(\vec r, \pvec r^{\prime})=\left. \frac{\delta^2 T_s[n]}{\delta n(\vec r)\delta n(\pvec r^{\prime})}\right|_{n=n_{\rm eq}}=-\left[\chi_{\rm KS}(\vec r, \pvec r^{\prime})\right]^{-1}.
\end{equation}

One can use the representation of relation (\ref{eq:EL2}) in Fourier space:
\begin{equation}\label{eq:EL3}
    K_{\rm s}(\vec q, \pvec q^{\prime})=\mathcal{F}\left[ \left. \frac{\delta^2 T_s[n]}{\delta n(\vec r)\delta n(\pvec r^{\prime})}\right|_{n=n_{\rm eq}} \right]=-\left[\chi_{\rm KS}(\vec q, \pvec q^{\prime})\right]^{-1},
\end{equation}
where $\mathcal{F}$ denotes the double spatial Fourier transform operation.

From Eq. (\ref{eq:EL2}) and Eq. (\ref{eq:EL3}), it follows that linear response theory allows one to compute the second-order functional derivative of the KE functional. 
This relation is the key element for the construction of the nonlocal KE functionals \cite{Wenhui_ChemRev_2023, Qiang_WIRE_2024, Huang_Carter_prb_2010, Mi_jcp_2018, Moldabekov_prb_2023}.

\subsection{Statement of the problem}
In OF-DFT simulations of materials, the use of pseudopotentials is crucial for the simulation's computational efficiency as well as accuracy \cite{Wenhui_ChemRev_2023}. The main idea is that by pseudizing the core electrons, the resulting valence electron densities are smooth and more closely resembling the density of a uniform system compared to the non-pseudized system. Resemblance to the UEG means that the employment of KE functionals based on the UEG will naturally result to improved accuracy.

In KS-DFT, the pseudization leads to pseudopotentials nonlocal in character (NLPs). This means, the resulting KS Hamiltonian is as follows
\begin{equation}
    \label{eq:KS_nl}
    \hat h_{nl} = -\frac{1}{2}\nabla^2 + v_H(\vec r) + v_{xc}(\vec r) + v_{loc}^{nl}(\vec r) + \hat v_{nl}
\end{equation}
where, $\hat v_{nl}$ is the nonlocal part of the NLP and $v_{loc}^{nl}(\vec r)$ is the local part of the NLP. This Hamiltonian differs from one which only considers LPs, $\hat h_{l}$, which will contain a local part of the pseudopotential, $v_{loc}^{l}(\vec r)\neq v_{loc}^{nl}(\vec r)$. This naturally will lead to different set of KS orbitals, $\{|\phi_i\rangle\}$, or KS density matrix, $\hat \gamma_s = \sum_i n_i |\phi_i\rangle\langle\phi_i|$. Thus, we have the following formal equations
\begin{eqnarray}
    \label{eq:KS_eql}
    \hat h_{l} | \phi^l_i \rangle &=& \varepsilon^l_i  | \phi^l_i \rangle\\
    \label{eq:KS_eqnl}
    \hat h_{nl} | \phi^{nl}_i \rangle &=& \varepsilon^{nl}_i  | \phi^{nl}_i \rangle
\end{eqnarray}
which lead to two sets of KS orbitals that crucially lead to the same density,
\begin{eqnarray}
    \label{eq:KS_denl}
    n(\vec r) = \sum_i n_i \langle \vec r |\phi_i^{nl}\rangle\langle\phi_i^{nl}|\vec r \rangle\\
    \label{eq:KS_dennl}
    n(\vec r) = \sum_i n_i \langle \vec r |\phi_i^{l}\rangle\langle\phi_i^{l}|\vec r \rangle\\
\end{eqnarray}
We note that the KS density matrices are generally not equal, $\hat \gamma_s^{nl}\neq \hat\gamma_s^{l}$. ($n_i$ are the orbital occupation numbers.)

We thus expect the KE functional to necessarily be different when using NLPs vs LPs. This is an aspect of OF-DFT simulations that is never considered. We thus analyze it in this work.

\section{Kernels via the external harmonic perturbation method}\label{sec:theory2}
\subsection{Evaluation of response functions in reciprocal space}\label{ss:method}
We consider a system in a cubic cell of length $L$ with periodic boundary conditions and 
 compute the response of this system to an external harmonic perturbation of the form:
\begin{equation}\label{eq:vext_r}
    \Delta v_{\rm ext}(\vec r)=2A\cos(\vec r \cdot\vec q_{\rm ext}),
\end{equation}
where $A$ and $ \vec q_{\rm ext}$ are the magnitude and the wavenumber of the perturbation. We set $ \vec q_{\rm ext}$ equal to a reciprocal lattice vector, e.g., choosing a perturbation along the $x$-axis, we have $q_{\rm ext}=j\times 2\pi/L$ with $j$ being an integer number. 

The Fourier transform of $v_{\rm ext}(\vec r)$ reads
\begin{equation}
     \Delta v_{\rm ext}(\vec q_j)=A\left[\delta_{\vec q_j, \vec q_{\rm ext}}+\delta_{\vec q_j, -\vec q_{\rm ext}}\right].
\end{equation}

In the linear response regime, the static density response $\Delta n(\vec q_j)$ at $\vec q_j$ to the perturbation  $ \Delta v_{\rm ext}(\vec q_i)$
is defined by the density response function:
\begin{eqnarray}\label{eq:chi_qqext}  
    \Delta n(\vec q_j)&=\sum_i \chi\left(\vec q_j,\vec q_i\right)\Delta v_{\rm ext}(\vec q_j) \label{eq:chi_qqext0} \\
    &=A\left[\chi\left(\vec q_j,\vec q_{\rm ext}\right)+\chi\left(\vec q_j,-\vec q_{\rm ext}\right)\right]\label{eq:chi_qqext} 
\end{eqnarray}

We consider the response of the system at $\vec q=\vec q_{\rm ext}$ and take into account that $\chi(\vec q_{\rm ext}, \vec q_{\rm ext})=\chi(\vec q_{\rm ext}, -\vec q_{\rm ext})$ in the bulk of a material, which means that the static response of the system at  $\vec q=\vec q_{\rm ext}$ is symmetric with respect to the perturbation in  $\vec q_{\rm ext}$ and $-\vec q_{\rm ext}$ directions. As the result, from Eq. (\ref{eq:chi_qqext}) it follows that:
\begin{equation}
\hspace*{-1cm}  
    \Delta n(\vec q_{\rm ext})=2A\chi\left(\vec q_{\rm ext},\vec q_{\rm ext}\right)
\end{equation}
and
\begin{equation}\label{eq:chi_qextqext}
\hspace*{-1cm}  
     \chi\left(\vec q_{\rm ext}\right)=\chi\left(\vec q_{\rm ext},\vec q_{\rm ext}\right)=\Delta n(\vec q_{\rm ext})/\left(2A\right)\ .
\end{equation}

Taking into account the cosine shape of the external perturbation $\Delta v_{\rm ext}(\vec r)$, the KS potential in the linear response regime is given by~\cite{Moldabekov_JCP_averaging}:
\begin{eqnarray} 
\hspace*{-1cm}  
    \Delta v_{\rm KS}(\vec r)&=2\sum_i u_{\rm KS}(\vec q_i) \cos\left(\vec q_i\, \vec r\right)\label{eq:uks_all}\\
\hspace*{-1cm}  
    &=2u_{\rm KS}(\vec q_{\rm ext}) \cos\left(\vec q_{\rm ext}\, \vec r\right)+\left[ \Delta v_{\rm KS}(\vec r)\right]^{(2)},
\end{eqnarray}
where we introduced the decomposition into the perturbation component at $\vec q_{\rm ext}$ and the contributions from higher harmonics~\cite{Dornheim_PRR_2021}:  
\begin{eqnarray}\label{eq:ks_cond}
    \left[ \Delta v_{\rm KS}(\vec r)\right]^{(2)}=2\sum_{q_i\neq q_{\rm ext}} u_{\rm KS}(\vec q_i) \cos\left(\vec q_i\, \vec r\right),
\end{eqnarray}
with $u_{\rm KS}(\vec q_i) $ being the Fourier component of the KS potential perturbation at $\vec q_i=i \vec q_{\rm ext}$. The factor two in Eq. (\ref{eq:ks_cond}) is conventional [cf. the perturbation term in Eq. (\ref{eq:vext_r})].

If the following condition is satisfied: 
\begin{eqnarray}\label{eq:condition}
    \left|\frac{\left[ \Delta v_{\rm KS}(\vec r)\right]^{(2)}}{\Delta v_{\rm KS}(\vec r, \vec q_{\rm ext})}\right|\ll 1,
\end{eqnarray}
one can compute the perturbation in the KS potential with good accuracy using
\begin{eqnarray}\label{eq:v_ks_r_cos}
    \Delta v_{\rm KS}(\vec r)\simeq 2u_{\rm KS}(\vec q_{\rm ext}) \cos\left(\vec q_{\rm ext}\, \vec r\right).
\end{eqnarray}

The Fourier transform of Eq. (\ref{eq:v_ks_r_cos}) at $\vec q=\vec q_j$ readily follows:
\begin{eqnarray}\label{eq:v_ks_Q}
    \Delta v_{\rm KS}(\vec q_{j})= u_{\rm KS}(\vec q_{\rm ext}) \left[\delta_{\vec q_j, \vec q_{\rm ext}}+\delta_{\vec q_j, -\vec q_{\rm ext}}\right].
\end{eqnarray}

Now, we use the Fourier transform of Eq. (\ref{eq:chiks}) for $\vec q=\vec q_{i}$ to find:
\begin{eqnarray}
    \Delta n(\vec q_j)&=\sum_i \chi_{\rm KS}\left(\vec q_j,\vec q_i\right)\Delta v_{\rm KS}(\vec q_i)\nonumber\\
    &= u_{\rm KS}(\vec q_{\rm ext})\left[\chi_{\rm KS}\left(\vec q_j,\vec q_{\rm ext}\right)+\chi_{\rm KS}\left(\vec q_j,-\vec q_{\rm ext}\right)\right].\label{eq:chi_ks_qqext} 
\end{eqnarray}

Finally, from Eq. (\ref{eq:chi_ks_qqext}), we deduce a relation for computing the KS response function at $\vec q=\vec q_{\rm ext}$:
\begin{eqnarray}\label{eq:chi_ks_fin}
    \chi_{\rm KS}\left(\vec q_{\rm ext}\right)=\chi_{\rm KS}\left(\vec q_{\rm ext},\vec q_{\rm ext}\right)=\frac{\Delta n(\vec q_{\rm ext})}{2u_{\rm KS}(\vec q_{\rm ext})},
\end{eqnarray}
where $u_{\rm KS}(\vec q_{\rm ext})$ is computed according to Eq. (\ref{eq:v_ks_r_cos}) using $\Delta v_{\rm KS}(\vec r, \vec q_{\rm ext})$ from KSDFT or OFDFT, and $\Delta n(\vec q_{\rm ext})$ is also directly extracted from the DFT simulations.

To sum up, one can compute the density response and KS response functions by computing the change in density and KS potential induced by harmonic external perturbation (\ref{eq:vext_r}) at different values of $\vec q=\vec q_{\rm ext}$, which have to be commensurate with the simulation cell periodicity. 
In this method, the accuracy of the calculation of the KS response function depends on what degree condition (\ref{eq:ks_cond}) is satisfied. This is not an uncontrolled approximation since the ratio on the left-hand side of Eq. (\ref{eq:ks_cond}) can be checked for every simulation [we demonstrate that in Sec. \ref{sec:results}]. 
The main criterion for the described method is thus that the perturbation magnitude $A$ is small such that linear response theory is applicable to describe the perturbations  $\Delta n(\vec r)$ and  $\Delta v_{\rm KS}(\vec r)$ \cite{Moldabekov_JCTC_2023, Dornheim_PRR_2021, Dornheim_PRL_2020, Bohme_PRL_2022}. 

\subsection{Relation to selected kernels of nonlocal (two-point) functionals}
For the analysis of the KSDFT simulation results, we use two popular approximations to the KE kernel.
The first is based on the UEG model, which is appropriate for the description of metals, and the second is based on the UEG-with-gap model, which is a more suitable approximation for semiconductors. 

\subsubsection{UEG model based KE kernel.}
\leavevmode\newline
To find the KE functional using information about the KS response function, the following standard decomposition is often used~\cite{Huang_Carter_prb_2010, Sjostrom_prb_2013, Wenhui_ChemRev_2023}:
\begin{equation}\label{eq:Esw}
    T_s[n]= T_{\rm NL}[n]+T_{\rm TF}[n]+T_{\rm vW}[n],
\end{equation}
where $T_{\rm NL}[n]$ is the nonlocal part of the KE functional, and $T_{\rm TF}[n]$ and $T_{\rm vW}[n]$ are the Thomas-Fermi (exact high-density limit) and the von Weizs\"acker (exact single-particle limit) KE functionals. 

A widely used efficacious approximation for the non-local part of the KE functional was introduced by Wang and Teter (WT)~\cite{WT_prb_1992}:
\begin{equation}\label{eq:KE_WT} 
    E_{\rm NL}[n]\simeq E_{\rm WT}[n]= \int n^{a}(\vec r) w(\vec r- \pvec r^{\prime}; n_0) n^{b}(\vec r) \mathrm{d}\vec r \mathrm{d}\pvec r^{\prime}
\end{equation}
where $a$ and $b$ can be chosen by trial and error to achieve an optimal result in terms of accuracy and transferability.
In Eq. (\ref{eq:KE_WT}),  $w(\vec r- \pvec r^{\prime})$ is approximated by the KE kernel computed using the density response function of an ideal uniform electron gas (Lindhard function) at the mean value of the valence electrons density \cite{Sjostrom_prb_2013}:
\begin{eqnarray}\label{eq:w_rr1}
\hspace*{-1cm} 
\mathcal{F}\left[ \left. \frac{\delta^2 E_{\rm WT}[n]}{\delta n(\vec r)\delta n(\pvec r^{\prime})}\right|_{n=n_{0}} \right]\nonumber&=w(\vec q; n_0)\times 2abn_0^{a+b-2}\\
 \hspace*{-1cm} 
 &=-\frac{1}{\chi_{\rm Lin}(q;n_0)}+\frac{1}{\chi_{\rm TF}(n_0)}+\frac{1}{\chi_{\rm vW}(q;n_0)},
 \label{eq:Kwt}
\end{eqnarray}
where $\chi_{\rm Lin}(q;n_0)^{-1}$, $\chi_{\rm TF}^{-1}=-\pi^2/q_F$ and $\chi_{\rm vW}^{-1}=-3\pi^2/q_F\times q^2/(2q_F)^2$ are computed using the Fermi wavenumber of the UEG at density $n_0=N/V$ with $N$ being the number of the valence electrons in the simulation cell of volume $V=L^3$.


\subsubsection{UEG-with-gap model based KE kernel.}
\leavevmode\newline
For semiconductors, the UEG-with-gap model \cite{Levine_prb_1982} is useful for constructing KE kernels \cite{Constantin_prb_2017, Bhattacharjee_2024} and for performing a qualitative analysis of the KS response function \cite{Huang_Carter_prb_2010}.
As an analogy of the Lindhard function of the UEG model, the ideal density response of the UEG-with-gap model reads \cite{Levine_prb_1982, Constantin_prb_2017}:
\begin{eqnarray}
\hspace*{-1cm} 
   \chi_{\rm gap}(q;n_0)&=-\frac{q_F}{\pi^2}\left[\frac{1}{2}-\frac{\Delta_{\rm gap}}{8z}{\left(\arctan\left(\frac{4z+z^2}{\Delta_{\rm gap}}\right)+\arctan\left(\frac{4z-z^2}{\Delta_{\rm gap}}\right)\right)} \right.\\
    &\left.+\left(\frac{\Delta_{\rm gap}^2}{128z^3}+\frac{1}{8z}-\frac{z}{8}\right)\ln\left(\frac{z^2+(4z+4z^2)^2}{z^2+(4z-4z^2)^2}\right)\right],
\end{eqnarray}
where $z=q/(2q_F)$ and $\Delta_{\rm gap}=E_{\rm gap}/E_F$ is the gap parameter defined as the ratio of the direct gap energy to Fermi energy. 

In contrast to the Lindhard function that has a non-zero constant limit $\lim\limits_{q\to0}\chi_{\rm Lin}(q;n_0)\to \chi_{\rm TF}(n_0)$, a peculiar property of $\chi_{\rm gap}(q; n_0)$ is that it tends to zero in the long wavelength limit as $\lim\limits_{q\to0}\chi_{\rm gap}(q;n_0)\to -q^2$. 
This property was used to construct KE functionals for semiconductors  \cite{Huang_Carter_prb_2010, Constantin_prb_2017, Xuecheng_prb_2021}. 
Among these functionals, the KE functional by Huang and Carter (HC) is arguably the most popular for OFDFT-based simulations of semiconductors \cite{Huang_Carter_prb_2010}.

We note that one can compute the density response function using $\chi_{\rm Lin}(q;n_0)$ or $\chi_{\rm gap}(q;n_0)$ as an ideal density response $\chi_0(q;n_0)$ in \cite{quantum_theory}: 
\begin{eqnarray}\label{eq:chi_ueg}
    \chi(q;n_0)=\frac{\chi_0(q;n_0)}{1-\left(\frac{4\pi^2}{q^2}+K_{\rm xc}(q;n_0)\right)\chi_0(q; n_0)},
\end{eqnarray}
where $K_{\rm xc}(q;n_0)$ is the exchange-correlation (XC) kernel defined as the second-order functional derivative of the XC functional $E_{\rm xc}[n]$ \cite{book_Ullrich}.
In this work, we use the XC functional in local density approximation (LDA) \cite{PhysRevB.45.13244}. 
Therefore, to compare $\chi(q;n_0)$ from the UEG and UEG-with-gap models with the KSDFT results, we use $K_{\rm xc}(q\to0;n_0)$ in the long wavelength approximation, where $K_{\rm xc}(q\to0)=\partial^2 E_{\rm xc}[n_0]/\partial n_0^2 $ follows from the compressibility sum rule \cite{PhysRevB.103.165102}.

Taking into account that it is a standard practice within the OFDFT community to use the decomposition (\ref{eq:Esw}) to represent and construct KE functionals \cite{Wenhui_ChemRev_2023, Qiang_WIRE_2024}, we analyze the KSDFT data for the KE kernel by considering the renormalization:

\begin{eqnarray}\label{eq:w_rr2}
\hspace*{-1cm} 
 \Delta K_{\rm s}(q)=K_{\rm s}(q)+\frac{1}{\chi_{\rm TF}(n_0)}+\frac{1}{\chi_{\rm vW}(q;n_0)},
 \label{eq:Kwt}
\end{eqnarray}
which, for free electrons, tends to zero at $q\to0$ (the TF limit) and to a constant at large wavenumbers (the vW limit). 

As shown in Sec. \ref{sec:results}, using $\Delta K_{\rm s}(q)$ gives one insights into the effects of
NLPs on the KE kernel at different length scales (wavenumbers).

\leavevmode\newline


\section{Simulation details}\label{sec:sim_details}

The KS-DFT calculations were performed using Quantum ESPRESSO  \cite{QE-2017, QE-2009, doi:10.1063/5.0005082}. To compute the perturbed density and KS potential, we used the QEpy tool \cite{qepy} to introduce a static external harmonic perturbation into the KSDFT simulations.
We consider bulk Al with a face-centered cubic (fcc) lattice with the lattice constant $a=4.05~{\rm \AA}$ and Si in the semiconducting crystal diamond state with  $a=5.408~{\rm \AA}$. The simulations of bulk Al are performed using cubic cells with $L=7.6534~{\rm \AA}$ containing 4 atoms and  $L=15.3068~{\rm \AA}$ containing 32 atoms. For bulk Si, we used  $L=10.2196~{\rm \AA}$ (with 8 atoms)  and $L=20.4393~{\rm \AA}$ (with 64 atoms).  All used pseudopotentials are from the Quantum ESPRESSO pseudopotential database. 
For Al, we used the pseudopotential from the projector augmented wave method {Al.pz-n-kjpaw\textunderscore psl.0.1.UPF} (PAW) and the ultrasoft pseudopotential {Al.pz-n-rrkjus\textunderscore psl.0.1.UPF} (USP). For Si, we used the PAW potential {Si.pz-n-kjpaw\textunderscore psl.0.1.UPF}  and the USP potential {Si.pz-n-kjpaw\textunderscore psl.0.1.UPF}. The kinetic energy cutoff for the wavefunctions is set to $E_{\rm cut}=74~{\rm Ry}$ ($E_{\rm cut}\simeq 1000 ~{\rm eV}$). The kinetic energy cutoff for the charge density and potential is set to $4E_{\rm cut}$. We used a $k$-point sampling of $12\times12\times12$.
The OFDFT simulations are performed using the DFTpy 
\cite{DFTpy} with bulk-derived local pseudopotentials (BLP) for Al and Si by Huang and Carter \cite{B810407G}. In all calculations, we take the XC functional in local density approximation \cite{PhysRevB.45.13244}.
The amplitude of the perturbation in Eq. (\ref{eq:vext_r}) is set to $A=0.01$ (in Hartree), which ensures that the induced density perturbation in both Al and Si is well within the linear response regime \cite{Moldabekov_prb_2023}.
The wavenumber of the perturbation is set along the $x$-axis with the absolute value defined by  $q=j\times q_{\rm min}$, where $q_{\rm min}=2\pi/L$ and $j$ is a positive integer number.

\begin{figure}\centering\includegraphics[width=1\textwidth]{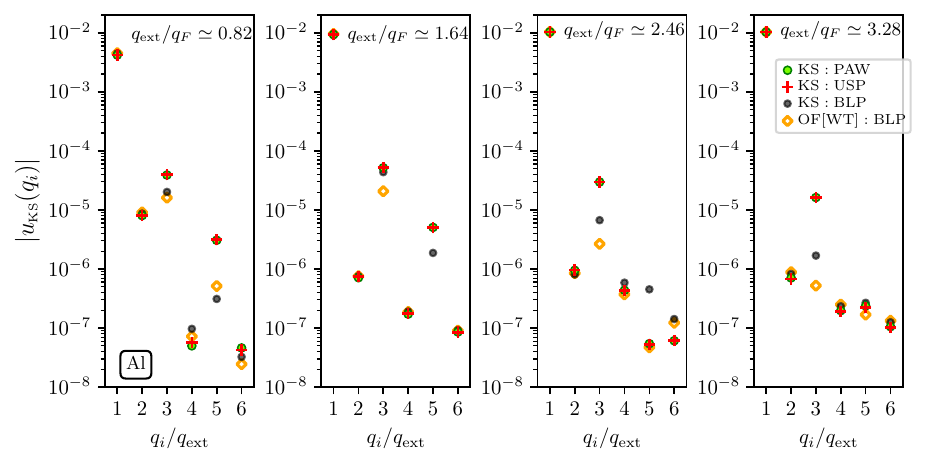}
\caption{\label{fig:Al_uG}
DFT results for the magnitudes of the perturbations of the KS potential at different wave numbers in the bulk of fcc Al. The external perturbation amplitude is set to  $A=0.01~{\rm Ha}$.
}
\end{figure}  

\section{Results\label{sec:results}}

To gain insight into the impact of NLPs onto the KE kernel, we examine fcc Al and cd Si, two typical examples of a simple metal and a semiconductor, respectively.
We compare the results for the KE kernel computed using PAW, USP, and BLP pseudopotentials in our KSDFT simulations.
To get a physical picture of the behavior of the KE kernel at different wavenumbers, we compare KSDFT results with both UEG and UEG-with-gap models.
Furthermore, we compare the KSDFT results with the OFDFT data computed using the WT functional for Al and the HC functional for Si.
This allows us to identify what kind of features of the KS kernel are missing in OFDFT simulations due to the application of local pseudopotentials. 

\subsection{Bulk aluminum\label{sec:Al_results}}

The applicability of the method presented in Sec. \ref{ss:method} requires the satisfaction of condition (\ref{eq:condition}).
Therefore, we need to first verify that it is indeed true in the considered case.
In Fig. \ref{fig:Al_uG},  we display the magnitude of $u_{\rm KS}(q_i)$ contributing to the total perturbation of the KS potential in accordance with Eq. (\ref{eq:uks_all})  at various values of the wavenumber of the external perturbation.
We find that the most significant contribution comes from $q_{i=1}=q_{\rm ext}$. 
This contribution is at least two orders of magnitude larger than the contributions at $q_{i>1}$ for all considered PPs and the wavenumbers of the external perturbation.

\begin{figure}\centering\includegraphics[width=1\textwidth]{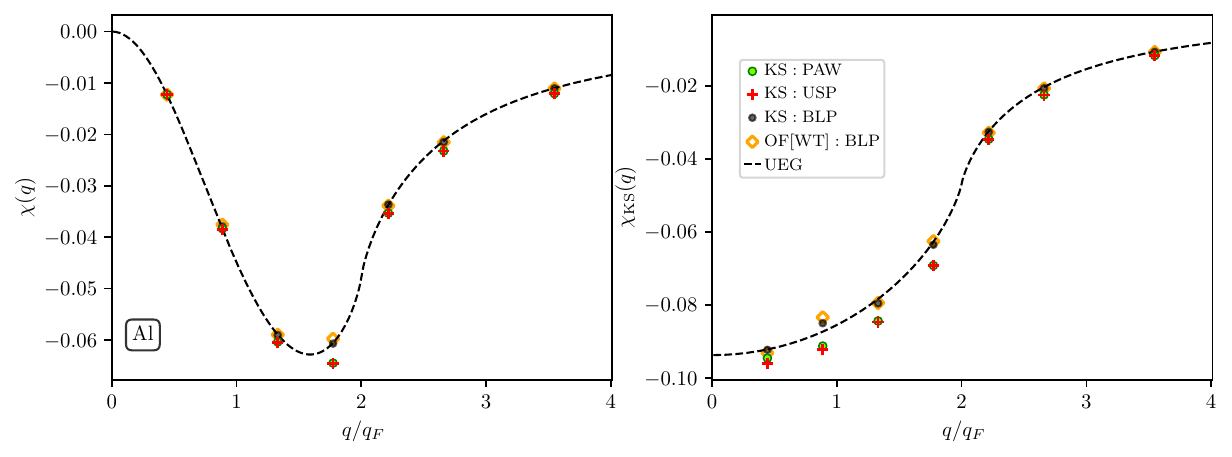}
\caption{\label{fig:Al_chi}
Left and right: Static density response function and KS response function of the valence electrons in fcc Al. 
DFT results are compared with UEG model calculations (dashed lines). 
}
\end{figure}

\begin{figure}\centering\includegraphics[width=0.7\textwidth]{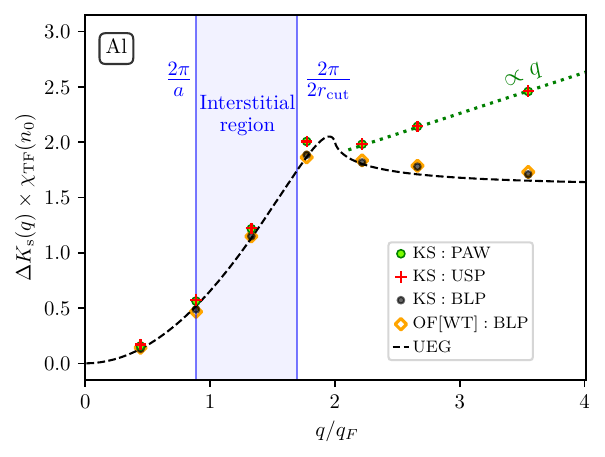}
\caption{\label{fig:Kin_kernel_Al}
The renormalized kinetic energy kernel [Eq.~(\ref{eq:w_rr2})] of the valence electrons in the bulk of fcc Al computed using different pseudopotentials. 
The DFT results are compared with the UEG model (dashed line). The vertical solid lines indicate the wavenumbers $2\pi/a$ (with $a$ being the lattice parameter) and $2\pi/(2r_{\rm cut})$ (with $2r_{\rm cut}$ being the diameter of the ion with core electrons in PAW). 
The perturbations with the wavelengths $a<\lambda<2r_{\rm cut}$  (wavenumbers $2\pi/a<q<2\pi/(2r_{\rm cut})$) probe the kinetic energy kernel of the electrons in the interstitial region.
}
\end{figure}  

Setting $q=q_{\rm ext}$, the results for the density response $\chi(q)$ and KS response $\chi_{\rm KS}(q)$ functions follow from Eq. (\ref{eq:chi_qextqext}) and Eq. (\ref{eq:chi_ks_fin}), respectively.
The corresponding results for $\chi(q)$ and $\chi_{\rm KS}(q)$ are shown in the left and right panels of Fig. \ref{fig:Al_chi}, and compared with the UEG model (dashed line) evaluated at the density parameter of the valence electrons $r_s=2.0738$, where $r_s=(4\pi n_0/3)^{-1/3}$. 
%
%
Clearly, the static density response properties of the valence electrons in the bulk of fcc Al resemble that of the UEG, 
which is particularly the case for the KSDFT and OFDFT data computed using the local pseudopotential BLP.
Proper incorporation of the effect of core electrons using NLPs leads to a noticeable difference in $\chi(q)$ around $q\simeq 1.5 q_F$ and in $\chi_{\rm KS}(q)$ at $q\lesssim 2q_F$. We note that $\chi_{\rm KS}(q)$ is connected with $\chi(q)$ 
by a similar relation as Eq. (\ref{eq:chi_ueg}).
Therefore, the differences between UEG and KSDFT results in $\chi_{\rm KS}$ at $q\lesssim 2q_F$ are suppressed in $\chi(q)$  as the result of screening on the mean-field level.

The results for $\Delta K_{\rm s}(q)$ (in units of $\chi_{\rm TF}^{-1}(n_0)$) are presented in Fig. \ref{fig:Kin_kernel_Al}, where $\Delta K_{\rm s}(q)$ is given by Eq. (\ref{eq:w_rr2}). We compare the KSDFT data with the OFDFT data (computed using the WT KE functional and the BLP), and with the UEG model results (i.e., $\Delta K_{\rm s}(q)$ computed using the Lindhard function). The vertical solid lines in Fig. \ref{fig:Kin_kernel_Al} correspond to the wavenumbers $2\pi/a$ (with $a$ being the lattice parameter) and $2\pi/(2r_{\rm cut})$ (with $2r_{\rm cut}$ being the diameter of the ion with core electrons in PAW). At $q<2\pi/a$, the density perturbation wavelength is larger than the length scale of the unit cell. At  $2\pi/a<q<2\pi/(2r_{\rm cut})$, the density perturbation wavelength is smaller than the lattice parameter, but larger than the ion-core diameter. 
We conventionally denote this domain as the interstitial region. At $q>2\pi/(2r_{\rm cut})$, the length scale of the perturbation is smaller than the diameter of the ion core. 
At $q<2\pi/a$, KSDFT and OFDFT data are in close agreement with each other and with the UEG model. In the interstitial region, the KSDFT data computed using NLPs have slightly larger values compared to the BLP-based KSDFT and OFDFT results, and the UEG model still provides a rather accurate description of the KE kernel of the electrons in Al. At $q>2\pi/(2r_{\rm cut})$, the KSDFT and OFDFT data computed using the local pseudopotential tends to the constant vW limit similar to the $\Delta K_{\rm s}(q)$ data computed using the Lindhard function. As one might expect, the effect of the NLPs is significant at  $q>2\pi/(2r_{\rm cut})$, where  $\Delta K_{\rm s}(q)$ increases linearly with the increase in the wavenumber. Finally, we note that the KSDFT data computed using PAW and USP are in excellent agreement with each other.

\begin{figure}\centering\includegraphics[width=1\textwidth]{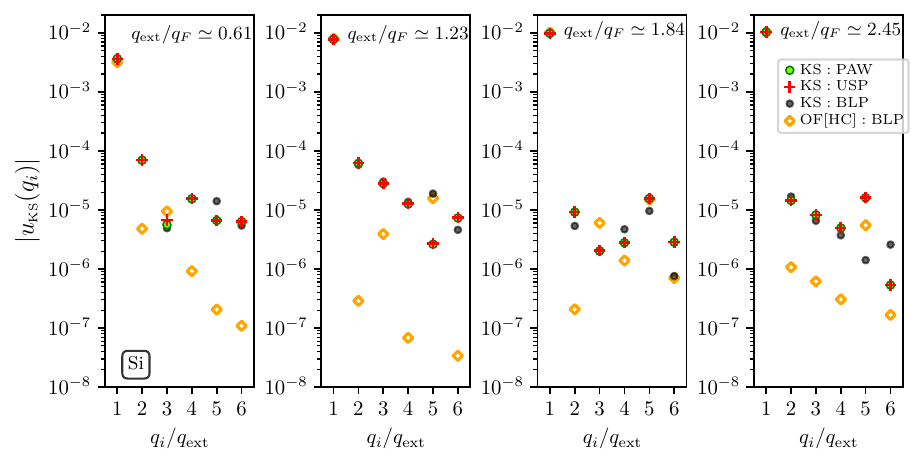}
\caption{\label{fig:Si_uG}
DFT results for the magnitudes of the perturbations of the KS potential at different wave numbers in the bulk of cd Si. 
The external perturbation amplitude is set to $A=0.01~{\rm Ha}$.
}
\end{figure}

\subsection{Bulk silicon\label{sec:Si_results}}

A less trivial example is given by Si in the semiconducting crystal diamond state. 
We checked the satisfaction of condition (\ref{eq:condition}) for the electrons in Si perturbed by harmonic external potentials with $A=0.01~{\rm Ha}$.
This is illustrated in Fig. \ref{fig:Si_uG}, where we observe that the KS potential perturbation is dominated by the term at $q_{\rm i=1}=q_{\rm ext}$ whereas terms corresponding to the perturbations $u_{\rm KS}(q_i)$ with $i>1$ are negligible.

\begin{figure}\centering\includegraphics[width=1\textwidth]{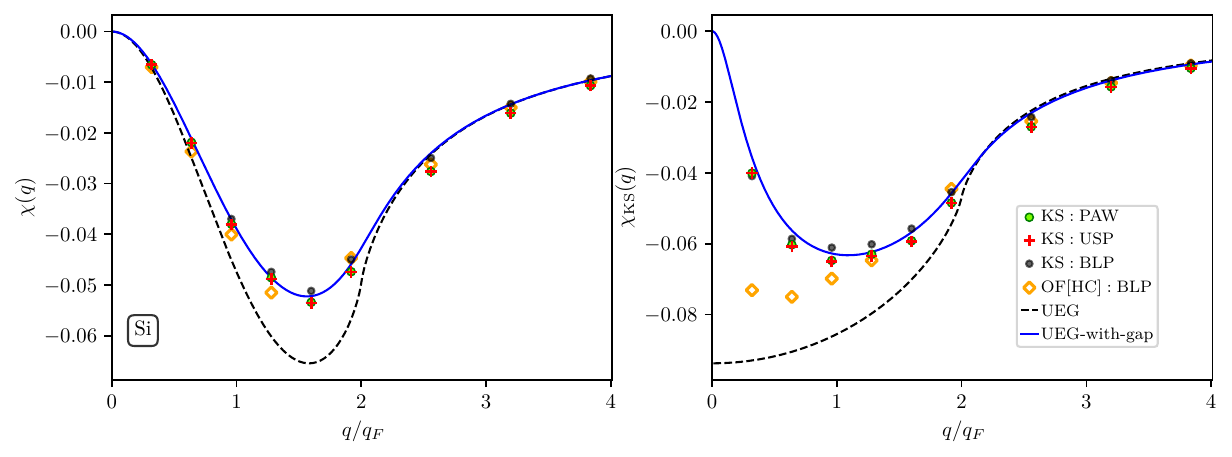}
\caption{\label{fig:Si_chi}
Left and right: Static density response function and KS response function 
of the valence electrons in cd Si.
The DFT results are compared with the calculations for the UEG model (dashed lines) and the UEG-with-gap model (solid lines). The energy gap parameter in the UEG-with-gap model is set to $\Delta=E_{\rm gap}/E_{F}=0.3968~{\rm eV}$.
}
\end{figure}

The results for the density response and KS response are shown in the left and right panels of Fig. \ref{fig:Si_chi}, respectively.
The OFDFT data were computed using the HC KE functional, which has been proven to provide adequate volumes and energies per unit cell as well as bulk moduli of Si semiconductor \cite{Huang_Carter_prb_2010}. We compare the DFT data with the UEG model (dashed line) and UEG-with-gap model (solid line) evaluated for the density parameter of the valence electrons $r_s=1.997$. In addition to $r_s$, the UEG-with-gap model requires $\Delta_{\rm gap}$ as an input parameter.
We used $\Delta=E_{\rm gap}/E_{F}=0.3968~{\rm eV}$ with the direct gap $E_{\rm gap}=2.532~{\rm eV}$ and the energy of the Fermi level $E_F=6.3808~{\rm eV}$, where $E_{\rm gap}$ and $E_F$ are from the KSDFT calculations using the PAW pseupotential. 
From Fig. \ref{fig:Si_chi}, first, we notice that for the density response function $\chi(q)$ (left panel of Fig. \ref{fig:Si_chi}), the difference between various simulation results as well as analytical models decreases with the decrease in the wavenumber at $q<q_F$.
The UEG-with-gap model has a decreasing trend with the limit $\lim\limits_{q\to0}\chi_{\rm gap}(q;n_0)\to -q^2$, which is enhanced by the Hartree mean field screening in Eq. (\ref{eq:chi_ueg}). In the UEG model, the Lindhard function tends to a constant at small wavenumbers, but the effect of screening also leads to the limit $\lim\limits_{q\to0}\chi(q;n_0)\to -q^2$.    At $q_F\lesssim q\lesssim 2q_F$, the difference between the UEG density response and the density response computed using the UEG-with-gap-model is the most substantial. On the other hand, the UEG-with-gap-model is in good agreement with the DFT results at these wavenumbers. 
With the increase in the perturbation wavenumber at $q>2q_F$, the density response of the system is reduced to such a degree that the difference between different presented results is not clearly visible at the considered scale.  

\begin{figure}\centering\includegraphics[width=0.7\textwidth]{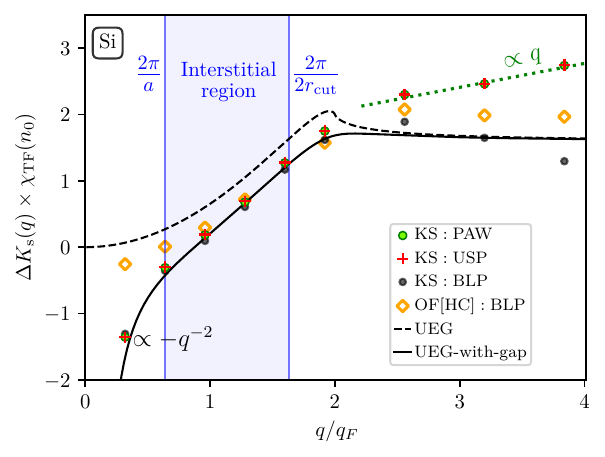}
\caption{\label{fig:Kin_kernel_Si}
Renormalized kinetic energy kernel [Eq.~(\ref{eq:w_rr2})] of the valence electrons in the bulk of cd Si computed using different pseudopotentials. 
The DFT results are compared with the UEG model calculations (dashed black line) and the UEG-with-gap model (solid black line).  The energy gap parameter in the UEG-with-gap model is set to $\Delta=E_{\rm gap}/E_{F}=0.3968~{\rm eV}$.
}
\end{figure}  

In the case of the KS response function, we see that the KSDFT results significantly disagree with the UEG model based results at $q<2q_F$ and with the OFDFT results at $q<q_F$, where the OFDFT data were computed using the HC KE functional. We notice that the disagreement with the OFDFT data at $q<q_F$ is substantially smaller than with the UEG model. In contrast, the UEG-with-gap model provides excellent agreement with the KSDFT data at $q<2q_F$.
At larger wavenumbers, $q>2q_F$, the disagreement between different calculation results for $\chi_{\rm KS}(q)$ is less pronounced than at $q<2q_F$.

The corresponding renormalized KE kernel $\Delta K_{\rm s}(q)$ is shown in Fig. \ref{fig:Kin_kernel_Si}, where we depict the values of wavevectors $2\pi/a$ and $2\pi/(2r_{\rm cut})$ by vertical lines.  
We find that at $q\lesssim 2\pi/(2r_{\rm cut})$, the UEG-with-gap model is in good agreement with the KSDFT data. 
Both the KSDFT results and the UEG-with-gap model attain negative values with a limit  $\Delta K_{\rm s}(q)\sim -q^2$ at $q<q_F$.
The OFDFT calculations are capable of correctly describing the KE kernel in the interstitial region with $2\pi/a< q \lesssim 2\pi/(2r_{\rm cut})$.
However, the OFDFT data show significant discrepancies compared to the KSDFT results at long wavelengths with  $ q \lesssim 2\pi/a$, although they are still somewhat better than the UEG model. As expected, the results computed using the Lindhard function (the UEG model) fail to correctly describe $\Delta K_{\rm s}(q)$  of Si in the semiconducting phase at all considered wavenumbers.
The KSDFT results computed using local (BLP) and nonlocal (PAW and USP) pseudopotentials are in close agreement at $q\lesssim 2\pi/(2r_{\rm cut})$ and start to deviate from each other with the increase in the wavenumber at  $q> 2\pi/(2r_{\rm cut})$. Therefore, for the KE kernel, the effect of using NLPs is pronounced for inhomogeneities with characteristic wavelengths smaller than the diameter of the ion core. With the increase in the wavenumber, for $q>2q_F$, we observe a linear increase of the $\Delta K_{\rm s}(q)$ values computed using NLPs. In contrast, the local pseudopotential-based KSDF results decrease with the increase in the wavenumber at $q>2q_F$. 
Interestingly, the large wavenumber limit of the OFDFT data differs from that of the UEG model, UEG-with-gap model, and the BLP-based KSDFT data. 
However, the OFDFT data tend to have a constant value at large $q$ similar to the UEG and UEG-with-gap models.

\section{Summary and Outlook\label{sec:summary}}

We have performed an analysis of the KE kernel focusing on the difference between the utilization of local and nonlocal pseudopotentials.
Specifically, we have introduced the direct harmonic perturbation approach, which gives access to both the static density response function and the KS response function.
The calculations are performed for metallic fcc Al and semiconducting Si in a diamond lattice. 
The KSDFT results are compared with the OFDFT data, with the  UEG model, and with the UEG-with-gap model.
We demonstrated that at $q<2\pi/(2r_{\rm cut})$, the UEG model and the UEG-with-gap model provide an excellent description of the KE kernel of  Al and Si, respectively.
For these materials, the effect of using NLPs is pronounced at $q>2\pi/(2r_{\rm cut})$, leading to a linear increase in $\Delta K_{\rm s}(q)$  with $q$. 
In contrast, we found that the bulk-derived pseudopotentials-based data for the KE kernel provide close agreement with the KSDFT results computed using NLPs in the interstitial region.

The presented methodology allows one to access the material-dependent KE kernel and, in this way, analyze various KE functionals on the level of the second-order functional derivatives. Since the KE kernel is the key ingredient to the construction of KE functionals, the presented approach can be useful for checking whether 
particular KE functionals show desirable behavior at different wavenumbers when applied to real materials. 
Furthermore, performing an analysis of the KE kernels of different materials using KSDFT might provide a valuable hint at a 
universal behavior of $K_{s}(q)$ at large wavenumbers, which can then be consistently incorporated into KE functionals for OFDFT applications. 
This can open up new avenues for the design of transferable KE functionals that properly take into account the effect of the core electrons. In the future, it should also be investigated to what degree such corrections are needed to improve the OFDFT simulations of equilibrium properties.  

Finally, we note that one important application of the OFDFT is simulating heated matter in the so-called warm dense matter (WDM) regime~\cite{wdm_book,new_POP,bonitz2024principles}, where the system temperature is comparable with the Fermi energy. 
At finite temperatures, the OFDFT method requires a non-interacting free energy functional $F_s[n]$ as an input. The second-order functional derivative of $F_s[n]$ with respect to the density is also equivalent to the inverse KS response function with a negative sign \cite{moldabekov2024review, Moldabekov_pop_2018} and a number of new advanced models for $F_s[n]$  have been recently developed, see Refs.~\cite{LKT_prb_2020, Ma_prb_2024, Karasiev_JPCL_2024, Sjostrom_prb_2013}.
One of the key methods of probing the microscopic properties of WDM in experiments is given by the X-ray Thomson scattering (XRTS) technique~\cite{siegfried_review}.
Analyzing XRTS data from a backward scattering measurements \cite{Tilo_Nature_2023,kraus_xrts} requires accurately simulating dynamic structure factors at large wave numbers $q\gg q_F$. Therefore, it is essential for the KE kernel to behave adequately at large wavenumbers when applying the OFDFT method to describe the XRTS spectrum at large scattering angles. The presented methodology can be particularly helpful in the development of $F_s[n]$ for such applications. 







\section*{Acknowledgments}
This work was partially supported by the Center for Advanced Systems Understanding (CASUS), financed by Germany’s Federal Ministry of Education and Research (BMBF) and the Saxon state government out of the State budget approved by the Saxon State Parliament. 
This work has received funding from the European Research Council (ERC) under the European Union’s Horizon 2022 research and innovation programme
(Grant agreement No. 101076233, "PREXTREME"). 
Views and opinions expressed are however those of the authors only and do not necessarily reflect those of the European Union or the European Research Council Executive Agency. Neither the European Union nor the granting authority can be held responsible for them. Computations were performed on a Bull Cluster at the Center for Information Services and High-Performance Computing (ZIH) at Technische Universit\"at Dresden and at the Norddeutscher Verbund f\"ur Hoch- und H\"ochstleistungsrechnen (HLRN) under grant mvp00024. 

\section*{References}

\bibliographystyle{unsrt}
\bibliography{bibliography}

\begin{thebibliography}{10}

\bibitem{MADDEN20069}
Paul~A. Madden, Robert Heaton, Andrés Aguado, and Sandro Jahn.
\newblock From first-principles to material properties.
\newblock {\em Journal of Molecular Structure: THEOCHEM}, 771(1):9--18, 2006.
\newblock Modelling Structure and Reactivity: the 7th triennial conference of the World Association of Theoritical and Computational Chemists (WATOC 2005).

\bibitem{Kohn_Sham}
W.~Kohn and L.~J. Sham.
\newblock Self-consistent equations including exchange and correlation effects.
\newblock {\em Phys. Rev.}, 140:A1133--A1138, Nov 1965.

\bibitem{Jones_RMP_2015}
R.~O. Jones.
\newblock Density functional theory: Its origins, rise to prominence, and future.
\newblock {\em Rev. Mod. Phys.}, 87:897--923, Aug 2015.

\bibitem{Unke2021}
Oliver~T. Unke, Stefan Chmiela, Huziel~E. Sauceda, Michael Gastegger, Igor Poltavsky, Kristof~T. Sch{\"u}tt, Alexandre Tkatchenko, and Klaus-Robert M{\"u}ller.
\newblock Machine learning force fields.
\newblock {\em Chemical Reviews}, 121(16):10142--10186, Aug 2021.

\bibitem{THOMPSON2022108171}
Aidan~P. Thompson, H.~Metin Aktulga, Richard Berger, Dan~S. Bolintineanu, W.~Michael Brown, Paul~S. Crozier, Pieter~J. {in 't Veld}, Axel Kohlmeyer, Stan~G. Moore, Trung~Dac Nguyen, Ray Shan, Mark~J. Stevens, Julien Tranchida, Christian Trott, and Steven~J. Plimpton.
\newblock Lammps - a flexible simulation tool for particle-based materials modeling at the atomic, meso, and continuum scales.
\newblock {\em Computer Physics Communications}, 271:108171, 2022.

\bibitem{Jiang_prb_2022}
Kaili Jiang, Xuecheng Shao, and Michele Pavanello.
\newblock Efficient time-dependent orbital-free density functional theory: Semilocal adiabatic response.
\newblock {\em Phys. Rev. B}, 106:115153, Sep 2022.

\bibitem{Krishtal_jcp_2015}
Alisa Krishtal, Davide Ceresoli, and Michele Pavanello.
\newblock {Subsystem real-time time dependent density functional theory}.
\newblock {\em The Journal of Chemical Physics}, 142(15):154116, 04 2015.

\bibitem{Gawne_prb_2024}
Thomas Gawne, Zhandos~A. Moldabekov, Oliver~S. Humphries, Karen Appel, Carsten Baehtz, Victorien Bouffetier, Erik Brambrink, Attila Cangi, Sebastian G\"ode, Zuzana Kon\^opkov\'a, Mikako Makita, Mikhail Mishchenko, Motoaki Nakatsutsumi, Kushal Ramakrishna, Lisa Randolph, Sebastian Schwalbe, Jan Vorberger, Lennart Wollenweber, Ulf Zastrau, Tobias Dornheim, and Thomas~R. Preston.
\newblock Ultrahigh resolution x-ray thomson scattering measurements at the european x-ray free electron laser.
\newblock {\em Phys. Rev. B}, 109:L241112, Jun 2024.

\bibitem{Moldabekov_PRR_2024}
Zhandos~A. Moldabekov, Thomas~D. Gawne, Sebastian Schwalbe, Thomas~R. Preston, Jan Vorberger, and Tobias Dornheim.
\newblock Excitation signatures of isochorically heated electrons in solids at finite wave number explored from first principles.
\newblock {\em Phys. Rev. Res.}, 6:023219, May 2024.

\bibitem{Moldabekov_ACSomega_2024}
Zhandos Moldabekov, Thomas~D. Gawne, Sebastian Schwalbe, Thomas~R. Preston, Jan Vorberger, and Tobias Dornheim.
\newblock Ultrafast heating-induced suppression of d-band dominance in the electronic excitation spectrum of cuprum.
\newblock {\em ACS Omega}, 9(23):25239--25250, 2024.

\bibitem{Wenhui_ChemRev_2023}
Wenhui Mi, Kai Luo, S.~B. Trickey, and Michele Pavanello.
\newblock Orbital-free density functional theory: An attractive electronic structure method for large-scale first-principles simulations.
\newblock {\em Chemical Reviews}, 123(21):12039--12104, 2023.
\newblock PMID: 37870767.

\bibitem{Qiang_WIRE_2024}
Qiang Xu, Cheng Ma, Wenhui Mi, Yanchao Wang, and Yanming Ma.
\newblock Recent advancements and challenges in orbital-free density functional theory.
\newblock {\em WIREs Computational Molecular Science}, 14(3):e1724, 2024.

\bibitem{PRR_2022}
Lenz Fiedler, Zhandos~A. Moldabekov, Xuecheng Shao, Kaili Jiang, Tobias Dornheim, Michele Pavanello, and Attila Cangi.
\newblock Accelerating equilibration in first-principles molecular dynamics with orbital-free density functional theory.
\newblock {\em Phys. Rev. Res.}, 4:043033, Oct 2022.

\bibitem{dftpy_paper}
Xuecheng Shao, Kaili Jiang, Wenhui Mi, Alessandro Genova, and Michele Pavanello.
\newblock Dftpy: An efficient and object-oriented platform for orbital-free dft simulations.
\newblock {\em WIREs Computational Molecular Science}, 11(1):e1482, 2021.

\bibitem{Dragon_paper}
Deyan~I. Mihaylov, S.X. Hu, and Valentin~V. Karasiev.
\newblock Dragon: A multi-gpu orbital-free density functional theory molecular dynamics simulation package for modeling of warm dense matter.
\newblock {\em Computer Physics Communications}, 294:108931, 2024.

\bibitem{Hamann_PRL_1979}
D.~R. Hamann, M.~Schl\"uter, and C.~Chiang.
\newblock Norm-conserving pseudopotentials.
\newblock {\em Phys. Rev. Lett.}, 43:1494--1497, Nov 1979.

\bibitem{Martin_2004}
Richard~M. Martin.
\newblock {\em Electronic Structure: Basic Theory and Practical Methods}.
\newblock Cambridge University Press, 2004.

\bibitem{Xu2022}
Qiang Xu, Cheng Ma, Wenhui Mi, Yanchao Wang, and Yanming Ma.
\newblock Nonlocal pseudopotential energy density functional for orbital-free density functional theory.
\newblock {\em Nature Communications}, 13(1):1385, Mar 2022.

\bibitem{B810407G}
Chen Huang and Emily~A. Carter.
\newblock Transferable local pseudopotentials for magnesium{,} aluminum and silicon.
\newblock {\em Phys. Chem. Chem. Phys.}, 10:7109--7120, 2008.

\bibitem{quantum_theory}
G.~Giuliani and G.~Vignale.
\newblock {\em Quantum Theory of the Electron Liquid}.
\newblock Cambridge University Press, Cambridge, 2008.

\bibitem{review}
T.~Dornheim, S.~Groth, and M.~Bonitz.
\newblock The uniform electron gas at warm dense matter conditions.
\newblock {\em Phys. Rep.}, 744:1--86, 2018.

\bibitem{loos}
P.-F. Loos and P.~M.~W. Gill.
\newblock The uniform electron gas.
\newblock {\em Comput. Mol. Sci}, 6:410--429, 2016.

\bibitem{Levine_prb_1982}
Zachary~H. Levine and Steven~G. Louie.
\newblock New model dielectric function and exchange-correlation potential for semiconductors and insulators.
\newblock {\em Phys. Rev. B}, 25:6310--6316, May 1982.

\bibitem{Constantin_prb_2017}
Lucian~A. Constantin, Eduardo Fabiano, Szymon \ifmmode~\acute{S}\else \'{S}\fi{}miga, and Fabio Della~Sala.
\newblock Jellium-with-gap model applied to semilocal kinetic functionals.
\newblock {\em Phys. Rev. B}, 95:115153, Mar 2017.

\bibitem{wdm_book}
F.~Graziani, M.~P. Desjarlais, R.~Redmer, and S.~B. Trickey, editors.
\newblock {\em Frontiers and Challenges in Warm Dense Matter}.
\newblock Springer, International Publishing, 2014.

\bibitem{Dornheim_review}
Tobias Dornheim, Zhandos~A. Moldabekov, Kushal Ramakrishna, Panagiotis Tolias, Andrew~D. Baczewski, Dominik Kraus, Thomas~R. Preston, David~A. Chapman, Maximilian~P. Böhme, Tilo Döppner, Frank Graziani, Michael Bonitz, Attila Cangi, and Jan Vorberger.
\newblock {Electronic density response of warm dense matter}.
\newblock {\em Phys. Plasmas}, 30(3):032705, 2023.

\bibitem{new_POP}
M.~Bonitz, T.~Dornheim, Zh.~A. Moldabekov, S.~Zhang, P.~Hamann, H.~Kählert, A.~Filinov, K.~Ramakrishna, and J.~Vorberger.
\newblock Ab initio simulation of warm dense matter.
\newblock {\em Phys. Plasmas}, 27(4):042710, 2020.

\bibitem{Wang_Carter_book}
Y.~A. Wang and E.~A. Carter.
\newblock {\em Orbital-Free Kinetic-Energy Density Functional Theory}, chapter~5, pages 117--184.
\newblock Kluwer, Dordrecht, 2000.

\bibitem{Huang_Carter_prb_2010}
Chen Huang and Emily~A. Carter.
\newblock Nonlocal orbital-free kinetic energy density functional for semiconductors.
\newblock {\em Phys. Rev. B}, 81:045206, Jan 2010.

\bibitem{Mi_jcp_2018}
Wenhui Mi, Alessandro Genova, and Michele Pavanello.
\newblock {Nonlocal kinetic energy functionals by functional integration}.
\newblock {\em The Journal of Chemical Physics}, 148(18):184107, 05 2018.

\bibitem{Moldabekov_prb_2023}
Zhandos~A. Moldabekov, Xuecheng Shao, Michele Pavanello, Jan Vorberger, Frank Graziani, and Tobias Dornheim.
\newblock Imposing correct jellium response is key to predict the density response by orbital-free dft.
\newblock {\em Phys. Rev. B}, 108:235168, Dec 2023.

\bibitem{Moldabekov_JCP_averaging}
Zhandos~A. Moldabekov, Jan Vorberger, Mani Lokamani, and Tobias Dornheim.
\newblock {Averaging over atom snapshots in linear-response TDDFT of disordered systems: A case study of warm dense hydrogen}.
\newblock {\em The Journal of Chemical Physics}, 159(1):014107, 07 2023.

\bibitem{Dornheim_PRR_2021}
Tobias Dornheim, Maximilian B\"ohme, Zhandos~A. Moldabekov, Jan Vorberger, and Michael Bonitz.
\newblock Density response of the warm dense electron gas beyond linear response theory: Excitation of harmonics.
\newblock {\em Phys. Rev. Research}, 3:033231, Sep 2021.

\bibitem{Moldabekov_JCTC_2023}
Zhandos Moldabekov, Maximilian B{\"o}hme, Jan Vorberger, David Blaschke, and Tobias Dornheim.
\newblock Ab initio static exchange--correlation kernel across jacob's ladder without functional derivatives.
\newblock {\em Journal of Chemical Theory and Computation}, 19(4):1286--1299, Feb 2023.

\bibitem{Dornheim_PRL_2020}
Tobias Dornheim, Jan Vorberger, and Michael Bonitz.
\newblock Nonlinear electronic density response in warm dense matter.
\newblock {\em Phys. Rev. Lett.}, 125:085001, Aug 2020.

\bibitem{Bohme_PRL_2022}
Maximilian B\"ohme, Zhandos~A. Moldabekov, Jan Vorberger, and Tobias Dornheim.
\newblock Static electronic density response of warm dense hydrogen: Ab initio path integral monte carlo simulations.
\newblock {\em Phys. Rev. Lett.}, 129:066402, Aug 2022.

\bibitem{Sjostrom_prb_2013}
Travis Sjostrom and J\'er\^ome Daligault.
\newblock Nonlocal orbital-free noninteracting free-energy functional for warm dense matter.
\newblock {\em Phys. Rev. B}, 88:195103, Nov 2013.

\bibitem{WT_prb_1992}
Lin-Wang Wang and Michael~P. Teter.
\newblock Kinetic-energy functional of the electron density.
\newblock {\em Phys. Rev. B}, 45:13196--13220, Jun 1992.

\bibitem{Bhattacharjee_2024}
Abhishek Bhattacharjee, Subrata Jana, and Prasanjit Samal.
\newblock {First step toward a parameter-free, nonlocal kinetic energy density functional for semiconductors and simple metals}.
\newblock {\em The Journal of Chemical Physics}, 160(22):224110, 06 2024.

\bibitem{Xuecheng_prb_2021}
Xuecheng Shao, Wenhui Mi, and Michele Pavanello.
\newblock Revised huang-carter nonlocal kinetic energy functional for semiconductors and their surfaces.
\newblock {\em Phys. Rev. B}, 104:045118, Jul 2021.

\bibitem{book_Ullrich}
Carsten~A. Ullrich.
\newblock {\em {Time-Dependent Density-Functional Theory: Concepts and Applications}}.
\newblock Oxford University Press, 12 2011.

\bibitem{PhysRevB.45.13244}
John~P. Perdew and Yue Wang.
\newblock Accurate and simple analytic representation of the electron-gas correlation energy.
\newblock {\em Phys. Rev. B}, 45:13244--13249, Jun 1992.

\bibitem{PhysRevB.103.165102}
Tobias Dornheim, Zhandos~A. Moldabekov, and Panagiotis Tolias.
\newblock Analytical representation of the local field correction of the uniform electron gas within the effective static approximation.
\newblock {\em Phys. Rev. B}, 103:165102, Apr 2021.

\bibitem{QE-2017}
P~Giannozzi, O~Andreussi, T~Brumme, O~Bunau, M~Buongiorno Nardelli, M~Calandra, R~Car, C~Cavazzoni, D~Ceresoli, M~Cococcioni, N~Colonna, I~Carnimeo, A~Dal Corso, S~de~Gironcoli, P~Delugas, R~A~DiStasio Jr, A~Ferretti, A~Floris, G~Fratesi, G~Fugallo, R~Gebauer, U~Gerstmann, F~Giustino, T~Gorni, J~Jia, M~Kawamura, H-Y Ko, A~Kokalj, E~Küçükbenli, M~Lazzeri, M~Marsili, N~Marzari, F~Mauri, N~L Nguyen, H-V Nguyen, A~Otero de-la Roza, L~Paulatto, S~Poncé, D~Rocca, R~Sabatini, B~Santra, M~Schlipf, A~P Seitsonen, A~Smogunov, I~Timrov, T~Thonhauser, P~Umari, N~Vast, X~Wu, and S~Baroni.
\newblock Advanced capabilities for materials modelling with quantum espresso.
\newblock {\em Journal of Physics: Condensed Matter}, 29(46):465901, 2017.

\bibitem{QE-2009}
Paolo Giannozzi, Stefano Baroni, Nicola Bonini, Matteo Calandra, Roberto Car, Carlo Cavazzoni, Davide Ceresoli, Guido~L Chiarotti, Matteo Cococcioni, Ismaila Dabo, Andrea {Dal Corso}, Stefano de~Gironcoli, Stefano Fabris, Guido Fratesi, Ralph Gebauer, Uwe Gerstmann, Christos Gougoussis, Anton Kokalj, Michele Lazzeri, Layla Martin-Samos, Nicola Marzari, Francesco Mauri, Riccardo Mazzarello, Stefano Paolini, Alfredo Pasquarello, Lorenzo Paulatto, Carlo Sbraccia, Sandro Scandolo, Gabriele Sclauzero, Ari~P Seitsonen, Alexander Smogunov, Paolo Umari, and Renata~M Wentzcovitch.
\newblock Quantum espresso: a modular and open-source software project for quantum simulations of materials.
\newblock {\em Journal of Physics: Condensed Matter}, 21(39):395502 (19pp), 2009.

\bibitem{doi:10.1063/5.0005082}
Paolo Giannozzi, Oscar Baseggio, Pietro Bonfà, Davide Brunato, Roberto Car, Ivan Carnimeo, Carlo Cavazzoni, Stefano de~Gironcoli, Pietro Delugas, Fabrizio Ferrari~Ruffino, Andrea Ferretti, Nicola Marzari, Iurii Timrov, Andrea Urru, and Stefano Baroni.
\newblock Quantum espresso toward the exascale.
\newblock {\em The Journal of Chemical Physics}, 152(15):154105, 2020.

\bibitem{qepy}
Xuecheng Shao, Oliviero Andreussi, Davide Ceresoli, Matthew Truscott, Andrew Baczewski, Quinn Campbell, and Michele Pavanello.
\newblock {QEpy: Quantum ESPRESSO in Python}.
\newblock {https://gitlab.com/shaoxc/qepy}.

\bibitem{DFTpy}
Xuecheng Shao, Kaili Jiang, Wenhui Mi, Alessandro Genova, and Michele Pavanello.
\newblock Dftpy: An efficient and object-oriented platform for orbital-free dft simulations.
\newblock {\em WIREs Computational Molecular Science}, 11(1):e1482, 2021.

\bibitem{bonitz2024principles}
Michael Bonitz, Jan Vorberger, Mandy Bethkenhagen, Maximilian Böhme, David Ceperley, Alexey Filinov, Thomas Gawne, Frank Graziani, Gianluca Gregori, Paul Hamann, Stephanie Hansen, Markus Holzmann, S.~X. Hu, Hanno Kählert, Valentin Karasiev, Uwe Kleinschmidt, Linda Kordts, Christopher Makait, Burkhard Militzer, Zhandos Moldabekov, Carlo Pierleoni, Martin Preising, Kushal Ramakrishna, Ronald Redmer, Sebastian Schwalbe, Pontus Svensson, and Tobias Dornheim.
\newblock First principles simulations of dense hydrogen, 2024.

\bibitem{moldabekov2024review}
Z.~Moldabekov, J.~Vorberger, and T.~Dornheim.
\newblock From density response to energy functionals and back: An ab initio perspective on matter under extreme conditions, 2024.

\bibitem{Moldabekov_pop_2018}
Zh.~A. Moldabekov, M.~Bonitz, and T.~S. Ramazanov.
\newblock {Theoretical foundations of quantum hydrodynamics for plasmas}.
\newblock {\em Physics of Plasmas}, 25(3):031903, 03 2018.

\bibitem{LKT_prb_2020}
K.~Luo, V.~V. Karasiev, and S.~B. Trickey.
\newblock Towards accurate orbital-free simulations: A generalized gradient approximation for the noninteracting free energy density functional.
\newblock {\em Phys. Rev. B}, 101:075116, Feb 2020.

\bibitem{Ma_prb_2024}
Cheng Ma, Min Chen, Yu~Xie, Qiang Xu, Wenhui Mi, Yanchao Wang, and Yanming Ma.
\newblock Nonlocal free-energy density functional for a broad range of warm dense matter simulations.
\newblock {\em Phys. Rev. B}, 110:085113, Aug 2024.

\bibitem{Karasiev_JPCL_2024}
Valentin~V. Karasiev, Joshua Hinz, and R.~M.~N. Goshadze.
\newblock Framework for laplacian-level noninteracting free-energy density functionals.
\newblock {\em The Journal of Physical Chemistry Letters}, 0(0):8272--8279, 0.
\newblock PMID: 39106051.

\bibitem{siegfried_review}
S.~H. Glenzer and R.~Redmer.
\newblock X-ray thomson scattering in high energy density plasmas.
\newblock {\em Rev. Mod. Phys}, 81:1625, 2009.

\bibitem{Tilo_Nature_2023}
T.~D{\"o}ppner, M.~Bethkenhagen, D.~Kraus, P.~Neumayer, D.~A. Chapman, B.~Bachmann, R.~A. Baggott, M.~P. B{\"o}hme, L.~Divol, R.~W. Falcone, L.~B. Fletcher, O.~L. Landen, M.~J. MacDonald, A.~M. Saunders, M.~Sch{\"o}rner, P.~A. Sterne, J.~Vorberger, B.~B.~L. Witte, A.~Yi, R.~Redmer, S.~H. Glenzer, and D.~O. Gericke.
\newblock Observing the onset of pressure-driven k-shell delocalization.
\newblock {\em Nature}, 618:270–275, May 2023.

\bibitem{kraus_xrts}
D.~Kraus, B.~Bachmann, B.~Barbrel, R.~W. Falcone, L.~B. Fletcher, S.~Frydrych, E.~J. Gamboa, M.~Gauthier, D.~O. Gericke, S.~H. Glenzer, S.~G\"ode, E.~Granados, N.~J. Hartley, J.~Helfrich, H.~J. Lee, B.~Nagler, A.~Ravasio, W.~Schumaker, J.~Vorberger, and T.~D\"oppner.
\newblock Characterizing the ionization potential depression in dense carbon plasmas with high-precision spectrally resolved x-ray scattering.
\newblock {\em Plasma Phys. Control Fusion}, 61:014015, 2019.

\end{thebibliography}

\end{document}